\documentstyle[prd,aps,preprint,tighten,epsfig]{revtex}
\begin{document}
\draft

\title{\Large\bf Unitarity Quadrangles of Four Neutrino Mixing}
\author{{\bf Wan-lei Guo} ~ and ~ {\bf Zhi-zhong Xing}}
\address{CCAST (World Laboratory), P.O. Box 8730, Beijing 100080, China; \\
and Institute of High Energy Physics, Chinese Academy of Sciences, \\
P.O. Box 918 (4), Beijing 100039, China 
\footnote{Mailing address} \\
({\it Electronic address: guowl@mail.ihep.ac.cn;
xingzz@mail.ihep.ac.cn})} \maketitle

\begin{abstract}
We present a classification of the unitarity quadrangles in the 
four-neutrino mixing scheme. We find that there are totally
thirty-six distinct topologies among twelve different unitarity
quadrangles. Concise relations are established between the areas 
of those unitarity quadrangles and the rephasing invariants of 
$CP$ and $T$ violation.
\end{abstract}


The robust Super-Kamiokande \cite{SK} and SNO \cite{SNO} data 
have provided convincing  evidence that the atmospheric $\nu_\mu$
neutrinos convert primarily into $\nu_\tau$ neutrinos, while
the solar $\nu_e$ neutrinos convert essentially into 
$\nu_\mu$ or $\nu_\tau$ neutrinos. It turns out that the 
existence of a light sterile neutrino $\nu_s$, which has been assumed
to reconcile the LSND \cite{LSND} evidence for 
$\overline{\nu}_\mu \rightarrow \overline{\nu}_e$ 
(and $\nu_\mu \rightarrow \nu_e$) oscillations 
with solar and atmospheric neutrino data in the four-neutrino
mixing scenarios \cite{4N}
\footnote{Instead of introducing a light sterile neutrino, 
a few more far-fetched ideas (such as the violation of $CPT$ symmetry
in the neutrino sector \cite{CPT} and the lepton-number-violating
muon decay \cite{Babu}) have been proposed in the literature.},
becomes questionable. Indeed,
a recent globable analysis of current neutrino oscillation data 
has shown that the well-known (2+2) and (3+1) schemes of
four neutrino mixing are both disfavored \cite{Valle}. The
upcoming MiniBooNE experiment \cite{MB} is therefore crucial, in order to 
confirm or disprove the LSND measurement. Before a definitely
negative conclusion can be drawn from MiniBooNE, however, the LSND data 
should be taken seriously. In particular, it is worthwhile to
investigate the four-neutrino mixing scenarios in a way 
without special theoretical biases and (or) empirical assumptions. 

In a recent paper \cite{GX}, we have calculated the rephasing invariants 
of $CP$ and $T$ violation by use of a favorable or ``standard''
parametrization of the 
generic $4\times 4$ neutrino mixing matrix. Our results
are expected to be quite useful for a systematical analysis of $CP$- and 
$T$-violating effects in various long-baseline neutrino oscillation
experiments. In the present work, which may serve as an important
addendum to Ref. \cite{GX}, we aim to present a complete geometrical
description of $CP$ and $T$ violation in the four-neutrino mixing scheme.

It is well known that the language of unitarity triangles is 
very helpful for the description of $CP$ violation in the 
quark sector \cite{PDG}. The same language has been introduced into
the lepton sector to describe $CP$ violation in the framework of
three-family lepton flavor mixing \cite{FX00,Branco,Sato,Smirnov}.
In Ref. \cite{Branco}, an example concerning the unitarity quadrangles
is given to illustrate the necessary condition of $CP$ violation 
in the four-neutrino mixing scenario
\footnote{The unitarity quadrangle discussed in Ref. \cite{Branco} is
equivalent to unitarity quadrangle ${\rm Q}_{e\mu}$ defined by us in
Eq. (3).}.
In this Brief Report, we shall first
make a classification of all possible unitarity quadrangles and then 
calculate their areas in terms of the rephasing invariants of $CP$ 
violation based on the standard parametrization of four neutrino mixing. 

Let us consider the admixture of three active
($\nu_e, \nu_\mu, \nu_\tau$) neutrinos and one sterile ($\nu_s$)
neutrino. Although $\nu_s$ does not participate in any normal 
weak interactions, it may oscillate with $\nu_e$, $\nu_\mu$ and 
$\nu_\tau$ \cite{SV}. Hence a $4\times 4$ unitary matrix $V$ is required
to fully describe four neutrino mixing in neutrino oscillations. In the
basis where the flavor and mass eigenstates of charged leptons are identical,
$V$ is defined to link the neutrino mass eigenstates
($\nu_0, \nu_1, \nu_2, \nu_3$) to the neutrino flavor 
eigenstates ($\nu_s, \nu_e, \nu_\mu, \nu_\tau$): 
\begin{equation}
\left ( \matrix{ \nu_s \cr \nu_e \cr \nu_\mu \cr \nu_\tau \cr} \right ) 
\; = \; \left ( \matrix{ V_{s0} & V_{s1} & V_{s2} &
V_{s3} \cr V_{e0} & V_{e1} & V_{e2} & V_{e3} \cr V_{\mu 0} &
V_{\mu 1} & V_{\mu 2} & V_{\mu 3} \cr V_{\tau 0} & V_{\tau 1} &
V_{\tau 2} & V_{\tau 3} \cr} \right ) \left ( \matrix{ \nu_0 \cr
\nu_1 \cr \nu_2 \cr \nu_3 \cr} \right ) \; .
\end{equation}
As neutrinos are expected to be Majorana particles, a full
parametrization of $V$ needs six mixing angles and six $CP$-violating
phases. Here we make use of the standard parametrization advocated
in Ref. \cite{Dai}; i.e.,
\small
\begin{equation}
V \; = \; \left ( \matrix{
c_{01}c_{02}c_{03} 
& c_{02}c_{03}\hat{s}_{01}^*
& c_{03}\hat{s}_{02}^* 
& \hat{s}_{03}^* 
\cr\cr
-c_{01}c_{02}\hat{s}_{03}\hat{s}_{13}^*
& -c_{02}\hat{s}_{01}^*\hat{s}_{03}\hat{s}_{13}^*
& -\hat{s}_{02}^*\hat{s}_{03}\hat{s}_{13}^*
& c_{03}\hat{s}_{13}^* 
\cr
-c_{01}c_{13}\hat{s}_{02}\hat{s}_{12}^*
& -c_{13}\hat{s}_{01}^*\hat{s}_{02}\hat{s}_{12}^*
& +c_{02}c_{13}\hat{s}_{12}^*
& 
\cr
-c_{12}c_{13}\hat{s}_{01}
& +c_{01}c_{12}c_{13}
&
&
\cr\cr
-c_{01}c_{02}c_{13}\hat{s}_{03}\hat{s}_{23}^*
& -c_{02}c_{13}\hat{s}_{01}^*\hat{s}_{03}\hat{s}_{23}^*
& -c_{13}\hat{s}_{02}^*\hat{s}_{03}\hat{s}_{23}^*
& c_{03}c_{13}\hat{s}_{23}^*
\cr
+c_{01}\hat{s}_{02}\hat{s}_{12}^*\hat{s}_{13}\hat{s}_{23}^*
& +\hat{s}_{01}^*\hat{s}_{02}\hat{s}_{12}^*\hat{s}_{13}\hat{s}_{23}^*
& -c_{02}\hat{s}_{12}^*\hat{s}_{13}\hat{s}_{23}^*
&
\cr
-c_{01}c_{12}c_{23}\hat{s}_{02}
& -c_{12}c_{23}\hat{s}_{01}^*\hat{s}_{02}
& +c_{02}c_{12}c_{23}
&
\cr
+c_{12}\hat{s}_{01}\hat{s}_{13}\hat{s}_{23}^*
& -c_{01}c_{12}\hat{s}_{13}\hat{s}_{23}^*
&
&
\cr 
+c_{23}\hat{s}_{01}\hat{s}_{12} 
& -c_{01}c_{23}\hat{s}_{12}
&
&
\cr\cr
-c_{01}c_{02}c_{13}c_{23}\hat{s}_{03}
& -c_{02}c_{13}c_{23}\hat{s}_{01}^*\hat{s}_{03}
& -c_{13}c_{23}\hat{s}_{02}^*\hat{s}_{03}
& c_{03}c_{13}c_{23}
\cr
+c_{01}c_{23}\hat{s}_{02}\hat{s}_{12}^*\hat{s}_{13}
& +c_{23}\hat{s}_{01}^*\hat{s}_{02}\hat{s}_{12}^*\hat{s}_{13}
& -c_{02}c_{23}\hat{s}_{12}^*\hat{s}_{13}
&
\cr
+c_{01}c_{12}\hat{s}_{02}\hat{s}_{23}
& +c_{12}\hat{s}_{01}^*\hat{s}_{02}\hat{s}_{23}
& -c_{02}c_{12}\hat{s}_{23}
&
\cr
+c_{12}c_{23}\hat{s}_{01}\hat{s}_{13}
& -c_{01}c_{12}c_{23}\hat{s}_{13}
&
&
\cr
-\hat{s}_{01}\hat{s}_{12}\hat{s}_{23}
& +c_{01}\hat{s}_{12}\hat{s}_{23}
&
&
\cr } \right ) \; ,
\end{equation}
\normalsize
where $c_{ij} \equiv \cos \theta_{ij}$ and
$\hat{s}_{ij} \equiv s_{ij} e^{{\rm i}\delta_{ij}}$ with
$s_{ij} \equiv \sin \theta_{ij}$. The strength of 
$CP$ and $T$ violation in neutrino oscillations 
is governed by the Jarlskog invariants 
$J^{ij}_{\alpha\beta} \equiv {\rm Im} ( V_{\alpha i}
V_{\beta j} V^*_{\alpha j} V^*_{\beta i} )$ \cite{J}, where
the Greek subscripts run over $(s, e, \mu, \tau)$ and the
Latin superscripts run over $(0, 1, 2, 3)$.

Note that a generic $n\times n$ lepton flavor mixing
matrix consists of $n (n-1)/2$ mixing angles, $(n-1) (n-2)/2$
Dirac-type phases and $(n-1)$ Majorana-type phases \cite{FX00,FX98}.
Furthermore, the number of independent Jarlskog invariants 
is $(n-1)^2 (n-2)^2/4$ \cite{Chau}. For $n=4$, we arrive at nine 
independent Jarlskog invariants compared to three Dirac-type $CP$-violating 
phases in $V$. This implies that there is no one-by-one correspondence
between Jarlskog parameters and Dirac-type phases, if the number of
lepton families is equal to or larger than four.
Given the parametrization of $V$ in Eq. (2) for four neutrino mixing, 
nine independent $J^{ij}_{\alpha\beta}$ are actually 
functions of three phase combinations (of the Dirac nature) and
six mixing angles, as explicitly shown in Ref. \cite{GX}. 
 
The unitarity of $V$ implies that there are twelve orthogonality
relations and eight normalization conditions among its sixteen matrix
elements. The former corresponds to twelve quadrangles in the 
complex plane, the so-called unitarity quadrangles. To be
explicit, let us write out the twelve orthogonality relations and
name their corresponding quadrangles:
\begin{eqnarray}
{\rm Q}_{se}: & ~ &
V_{s0}V_{e0}^{\ast}+V_{s1}V_{e1}^{\ast}+V_{s2}V_{e2}^{\ast}
+V_{s3}V_{e3}^{\ast}=0 \; ,
\nonumber \\
{\rm Q}_{s\mu}: & ~ &
V_{s0}V_{\mu0}^{\ast}+V_{s1}V_{\mu1}^{\ast}+V_{s2}V_{\mu2}^{\ast}
+V_{s3}V_{\mu3}^{\ast}=0 \; ,
\nonumber \\
{\rm Q}_{s\tau}: & ~ &
V_{s0}V_{\tau0}^{\ast}+V_{s1}V_{\tau1}^{\ast}+V_{s2}V_{\tau2}^{\ast}
+V_{s3}V_{\tau3}^{\ast}=0 \; ,
\nonumber \\ 
{\rm Q}_{e\mu}: & ~ &
V_{e0}V_{\mu0}^{\ast}+V_{e1}V_{\mu1}^{\ast}+V_{e2}V_{\mu2}^{\ast}
+V_{e3}V_{\mu3}^{\ast}=0 \; ,
\nonumber \\ 
{\rm Q}_{e\tau}: & ~ &
V_{e0}V_{\tau0}^{\ast}+V_{e1}V_{\tau1}^{\ast}+V_{e2}V_{\tau2}^{\ast}
+V_{e3}V_{\tau3}^{\ast}=0 \; ,
\nonumber \\
{\rm Q}_{\mu\tau}: & ~ &
V_{\mu0}V_{\tau0}^{\ast}+V_{\mu1}V_{\tau1}^{\ast}+V_{\mu2}V_{\tau2}^{\ast}
+V_{\mu3}V_{\tau3}^{\ast}=0 \; ;
\end{eqnarray}
and
\begin{eqnarray}
{\rm Q}_{01}: & ~ &
V_{s0}V_{s1}^{\ast}+V_{e0}V_{e1}^{\ast}+V_{\mu0}V_{\mu1}^{\ast}
+V_{\tau0}V_{\tau1}^{\ast}=0 \; ,
\nonumber \\
{\rm Q}_{02}: & ~ &
V_{s0}V_{s2}^{\ast}+V_{e0}V_{e2}^{\ast}+V_{\mu0}V_{\mu2}^{\ast}
+V_{\tau0}V_{\tau2}^{\ast}=0 \; ,
\nonumber \\ 
{\rm Q}_{03}: & ~ &
V_{s0}V_{s3}^{\ast}+V_{e0}V_{e3}^{\ast}+V_{\mu0}V_{\mu3}^{\ast}
+V_{\tau0}V_{\tau3}^{\ast}=0 \; ,
\nonumber \\ 
{\rm Q}_{12}: & ~ &
V_{s1}V_{s2}^{\ast}+V_{e1}V_{e2}^{\ast}+V_{\mu1}V_{\mu2}^{\ast}
+V_{\tau1}V_{\tau2}^{\ast}=0 \; ,
\nonumber \\
{\rm Q}_{13}: & ~ &
V_{s1}V_{s3}^{\ast}+V_{e1}V_{e3}^{\ast}+V_{\mu1}V_{\mu3}^{\ast}
+V_{\tau1}V_{\tau3}^{\ast}=0 \; ,
\nonumber \\
{\rm Q}_{23}: & ~ &
V_{s2}V_{s3}^{\ast}+V_{e2}V_{e3}^{\ast}+V_{\mu2}V_{\mu3}^{\ast}
+V_{\tau2}V_{\tau3}^{\ast}=0 \; .
\end{eqnarray}
If six mixing angles and six $CP$-violating phases of $V$ are all
known, one can plot twelve unitarity quadrangles without ambiguities.
Note, however, that each quadrangle has three distinct topologies
in the complex plane. For illustration, we take quadrangle ${\rm Q}_{se}$
for example and show its three topologies in FIG. 1, where 
the sizes and phases of $V_{si}V^*_{ei}$ (for $i=0,1,2,3$) have 
been fixed. One can see that different topologies of quadrangle
${\rm Q}_{se}$ arise from different orderings of its four sides,
and their areas are apparently different from one another. As a
whole, there are totally thirty-six different topologies among twelve 
unitarity quadrangles.

Now we calculate the areas of all unitarity triangles and
relate them to the rephasing invariants of $CP$ violation 
$J^{ij}_{\alpha\beta}$. Taking quadrangle ${\rm Q}_{se}$ as
an example again, we find that the areas of its three distinct 
topologies can be given respectively by
\footnote{It should be noted that the areas of unitarity quadrangles 
under discussion are ``algebraic areas'', namely, they can be either 
positive or negative. Of course, it is always possible to take 
$S^a_{se} = (|J^{10}_{se}|+|J^{21}_{se}|+|J^{32}_{se}|+|J^{03}_{se}|)/4$
or $S^a_{se} = |J^{10}_{se}+J^{21}_{se}+J^{32}_{se}+J^{03}_{se}|/4$,
such that $S^a_{se}$ is definitely positive.
We find, however, that the language of ``algebraic areas'' is simpler and
more convenient in the description of unitarty quadrangles.
In particular, the algebraic area of unitarity quadrangle ${\rm Q}_{se}$
in the case of topology (b) means an algebraic sum of the areas of
its two disassociated triangles, which have opposite signs. Hence 
both $S^b_{se}=0$ and $S^b_{se}<0$ are in general allowed. 
As for topologies (a) and (c) of ${\rm Q}_{se}$, $S^a_{se}>0$ and 
$S^c_{se}>0$ are simply a matter of sign or phase convention.}
\begin{eqnarray}
S^a_{se} & = & \frac{1}{4} \left [ 
{\rm Im} \left (V_{s1}V_{e0}V^*_{s0}V^*_{e1} \right ) +
{\rm Im} \left (V_{s2}V_{e1}V^*_{s1}V^*_{e2} \right ) +
{\rm Im} \left (V_{s3}V_{e2}V^*_{s2}V^*_{e3} \right ) +
{\rm Im} \left (V_{s0}V_{e3}V^*_{s3}V^*_{e0} \right ) \right ]
\nonumber \\
& = & \frac{1}{4} \left ( J^{10}_{se} + J^{21}_{se} + J^{32}_{se}
+ J^{03}_{se} \right ) \; ,
\nonumber \\
S^b_{se} & = & \frac{1}{4} \left [ 
{\rm Im} \left (V_{s2}V_{e0}V^*_{s0}V^*_{e2} \right ) +
{\rm Im} \left (V_{s1}V_{e2}V^*_{s2}V^*_{e1} \right ) +
{\rm Im} \left (V_{s3}V_{e1}V^*_{s1}V^*_{e3} \right ) +
{\rm Im} \left (V_{s0}V_{e3}V^*_{s3}V^*_{e0} \right ) \right ]
\nonumber \\
& = & \frac{1}{4} \left ( J^{20}_{se} + J^{12}_{se} + J^{31}_{se}
+ J^{03}_{se} \right ) \; ,
\nonumber \\
S^c_{se} & = & \frac{1}{4} \left [ 
{\rm Im} \left (V_{s1}V_{e0}V^*_{s0}V^*_{e1} \right ) +
{\rm Im} \left (V_{s3}V_{e1}V^*_{s1}V^*_{e3} \right ) +
{\rm Im} \left (V_{s2}V_{e3}V^*_{s3}V^*_{e2} \right ) +
{\rm Im} \left (V_{s0}V_{e2}V^*_{s2}V^*_{e0} \right ) \right ]
\nonumber \\
& = & \frac{1}{4} \left ( J^{10}_{se} + J^{31}_{se} + J^{23}_{se}
+ J^{02}_{se} \right ) \; ,
\end{eqnarray}
where $J^{ij}_{se}$ have been defined 
below Eq. (2). In a similar way, one may calculate the areas of
the other eleven unitarity quadrangles. The results for  
thirty-six different topologies of twelve unitarity quadrangles 
are summarized as follows:
\begin{eqnarray}
S_{\alpha\beta}^{a} & = & \frac{1}{4} \left (J_{\alpha\beta}^{10}
+J_{\alpha\beta}^{21}+J_{\alpha\beta}^{32}+J_{\alpha\beta}^{03} \right ) \; ,
\nonumber \\
S_{ij}^{a} & = & \frac{1}{4} \left (J_{es}^{ij}+J_{\mu
e}^{ij}+J_{\tau\mu}^{ij}+J_{s\tau}^{ij} \right ) \; ;
\end{eqnarray}
\begin{eqnarray}
S_{\alpha\beta}^{b} & = & \frac{1}{4} \left (J_{\alpha\beta}^{20}
+J_{\alpha\beta}^{12}+J_{\alpha\beta}^{31}+J_{\alpha\beta}^{03} \right ) \; ,
\nonumber \\
S_{ij}^{b} & = & \frac{1}{4} \left (J_{\mu
s}^{ij}+J_{e\mu }^{ij}+J_{\tau e}^{ij}+J_{s\tau}^{ij} \right ) \; ;
\end{eqnarray}
and
\begin{eqnarray}
S_{\alpha\beta}^{c} & = & \frac{1}{4} \left (J_{\alpha\beta}^{10}+
J_{\alpha\beta}^{31}+J_{\alpha\beta}^{23}+J_{\alpha\beta}^{02} \right ) \; ,
\nonumber \\
S_{ij}^{c} & = & \frac{1}{4} \left (J_{e
s}^{ij}+J_{\tau e}^{ij}+J_{\mu\tau}^{ij}+J_{s\mu}^{ij} \right ) \; ,
\end{eqnarray}
where the subscripts $\alpha\beta = se$, $s\mu$, $s\tau$,
$e\mu$, $e\tau$ or $\mu\tau$, and $ij = 01$, $02$, $03$, $12$, $13$ or $23$.
Note that the correlation of $J^{ij}_{\alpha\beta}$ \cite{GX,Dai} allows us to
simplify Eqs. (6), (7) and (8). Then each $S^q_{\alpha\beta}$ or 
$S^q_{ij}$ (for $q=a, b, c$) can be expressed as a sum of two independent 
Jarlskog invariants. Such simplified expressions of $S^q_{\alpha\beta}$
and $S^q_{ij}$ depend on the choice of independent $J^{ij}_{\alpha\beta}$,
therefore they may have many different forms. If nine independent
Jarlskog invariants are fixed, however, some expressions of 
$S^q_{\alpha\beta}$ and $S^q_{ij}$ must consist of three 
$J^{ij}_{\alpha\beta}$. This point will become clear later on.

Inversely, one may express $J^{ij}_{\alpha\beta}$ in terms of 
$S^q_{\alpha\beta}$ or $S^q_{ij}$. The explicit formulas are 
\begin{equation}
\left ( \matrix{J_{\alpha\beta}^{01} \cr J_{\alpha\beta}^{02} \cr
J_{\alpha\beta}^{03} \cr J_{\alpha\beta}^{12} \cr
J_{\alpha\beta}^{13} \cr J_{\alpha\beta}^{23} \cr} \right ) \; = \; 
\left ( \matrix{ -1 & 0 & -1  \cr 0 & -1 & 1  \cr 1 & 1 & 0 \cr
-1 & 1 & 0  \cr  0 & -1 & -1  \cr -1 & 0 & 1 \cr}  \right ) \left
( \matrix{S_{\alpha\beta}^{a} \cr\cr S_{\alpha\beta}^{b} 
\cr\cr S_{\alpha\beta}^{c} \cr} \right ) \; ,
\end{equation}
which corresponds to the unitarity quadrangles 
${\rm Q}_{se}$, ${\rm Q}_{s\mu}$,
${\rm Q}_{s\tau}$, ${\rm Q}_{e\mu}$, ${\rm Q}_{e\tau}$ 
and ${\rm Q}_{\mu\tau}$ defined in Eq. (3);
and
\begin{equation}
\left ( \matrix{J_{se}^{ij} \cr J_{s\mu}^{ij} \cr J_{s\tau}^{ij}
\cr J_{e\mu}^{ij} \cr J_{e\tau}^{ij} \cr J_{\mu\tau}^{ij} \cr}
\right ) \; = \; 
\left ( \matrix{ -1 & 0 & -1  \cr 0 & -1 & 1  \cr
1 & 1 & 0 \cr -1 & 1 & 0  \cr  0 & -1 & -1  \cr -1 & 0 & 1 \cr }
\right ) \left ( \matrix{ S_{ij}^{a} \cr\cr S_{ij}^{b} 
\cr\cr S_{ij}^{c} \cr} \right ) \; ,
\end{equation}
which corresponds to the unitarity quadrangles ${\rm Q}_{01}$, ${\rm Q}_{02}$,
${\rm Q}_{03}$, ${\rm Q}_{12}$, ${\rm Q}_{13}$ 
and ${\rm Q}_{23}$ defined in Eq. (4). 

As $J^{ij}_{\alpha\beta} = -J^{ji}_{\alpha\beta} =
-J^{ij}_{\beta\alpha} = J^{ji}_{\beta\alpha}$ holds by definition, 
one may easily obtain $S^q_{\alpha\beta} = -S^q_{\beta\alpha}$ and
$S^q_{ij} = -S^q_{ji}$ (for $q=a$, $b$, $c$). From 
the sum rule \cite{GX,Dai}
\begin{equation}
\sum_i J^{ij}_{\alpha\beta} \; = \; \sum_j J^{ij}_{\alpha\beta} \; = \;
\sum_\alpha J^{ij}_{\alpha\beta} \; = \; \sum_\beta
J^{ij}_{\alpha\beta} \; =\; 0 \; ,
\end{equation}
one can also find
\begin{equation}
\sum_\alpha S_{\alpha\beta}^q \; = \; \sum_\beta S_{\alpha\beta}^q \; = \;
\sum_i S_{ij}^q \; = \; \sum_j S_{ij}^q \; = \; 0 \; ,
\end{equation}
where $\alpha$ or $\beta$ runs over $(s, e, \mu, \tau)$, and $i$ or $j$
runs over $(0, 1, 2, 3)$. In addition to Eq. (12), the following
relations can be derived from Eqs. (6), (7) and (8):
\begin{eqnarray}
-S_{se}^{a}-S_{e\mu}^{a}-S_{\mu\tau}^{a}+S_{s\tau}^{a} & = &
-S_{01}^{a}-S_{12}^{a}-S_{23}^{a}+S_{03}^{a} \; ,
\nonumber \\
-S_{s\mu}^{a}+S_{e\mu}^{a}-S_{e\tau}^{a}+S_{s\tau}^{a} & = &
-S_{01}^{b}-S_{12}^{b}-S_{23}^{b}+S_{03}^{b} \; ,
\nonumber \\
-S_{se}^{a}-S_{e\tau}^{a}+S_{\mu\tau}^{a}+S_{s\mu}^{a} & = & 
-S_{01}^{c}-S_{12}^{c}-S_{23}^{c}+S_{03}^{c} \; ;
\end{eqnarray}
\begin{eqnarray}
-S_{se}^{b}-S_{e\mu}^{b}-S_{\mu\tau}^{b}+S_{s\tau}^{b} & = & 
-S_{02}^{a}+S_{12}^{a}-S_{13}^{a}+S_{03}^{a} \; ,
\nonumber \\
-S_{s\mu}^{b}+S_{e\mu}^{b}-S_{e\tau}^{b}+S_{s\tau}^{b} & = &
-S_{02}^{b}+S_{12}^{b}-S_{13}^{b}+S_{03}^{b} \; ,
\nonumber \\
-S_{se}^{b}-S_{e\tau}^{b}+S_{\mu\tau}^{b}+S_{s\mu}^{b} & = &
-S_{02}^{c}+S_{12}^{c}-S_{13}^{c}+S_{03}^{c} \; ;
\end{eqnarray}
and
\begin{eqnarray}
-S_{se}^{c}-S_{e\mu}^{c}-S_{\mu\tau}^{c}+S_{s\tau}^{c} & = &
-S_{01}^{a}-S_{13}^{a}+S_{23}^{a}+S_{02}^{a} \; ,
\nonumber \\
-S_{s\mu}^{c}+S_{e\mu}^{c}-S_{e\tau}^{c}+S_{s\tau}^{c} & = &
-S_{01}^{b}-S_{13}^{b}+S_{23}^{b}+S_{02}^{b} \; ,
\nonumber \\
-S_{se}^{c}-S_{e\tau}^{c}+S_{\mu\tau}^{c}+S_{s\mu}^{c} & = &
-S_{01}^{c}-S_{13}^{c}+S_{23}^{c}+S_{02}^{c} \; .
\end{eqnarray}
The correlation equations (12) -- (15) indicate that there are only
nine independent $S^q_{\alpha\beta}$ and (or) $S^q_{ij}$, corresponding
to nine independent $J^{ij}_{\alpha\beta}$.  

Without loss of generality, let us choose the following nine
independent $S^q_{\alpha\beta}$:
\begin{eqnarray}
S_{se}^{a} & = & \frac{1}{2} 
\left (J_{se}^{02} - J_{se}^{13} - 2J_{se}^{23} \right ) \; ,
\nonumber \\
S_{s\tau}^{a} & = & \frac{1}{2} 
\left (J_{s\tau}^{02} + J_{s\tau}^{03}- J_{s\tau}^{23} \right ) \; ,
\nonumber \\
S_{e\mu}^{a} & = & \frac{1}{2} 
\left (-J_{e\mu}^{12} - J_{e\mu}^{13}- J_{e\mu}^{23} \right ) \; ,
\nonumber \\
S_{se}^{b} & = & \frac{1}{2} 
\left (-J_{se}^{02} - J_{se}^{13} \right ) \; ,
\nonumber \\
S_{s\tau}^{b} & = & \frac{1}{2} 
\left (-J_{s\tau}^{02} + J_{s\tau}^{03}+ J_{s\tau}^{23} \right ) \; ,
\nonumber \\
S_{e\mu}^{b} & = & \frac{1}{2} 
\left (J_{e\mu}^{12} - J_{e\mu}^{13}- J_{e\mu}^{23} \right ) \; ,
\nonumber \\
S_{se}^{c} & = & \frac{1}{2} 
\left (J_{se}^{02} - J_{se}^{13} \right ) \; ,
\nonumber \\
S_{s\tau}^{c} & = & \frac{1}{2} 
\left (J_{s\tau}^{02} + J_{s\tau}^{03}+ J_{s\tau}^{23} \right ) \; ,
\nonumber \\
S_{e\mu}^{c} & = & \frac{1}{2} 
\left (-J_{e\mu}^{12} - J_{e\mu}^{13} + J_{e\mu}^{23} \right ) \; .
\end{eqnarray}
In terms of six flavor mixing angles and three independent phase 
combinations of $V$, we have expressed the nine independent 
$J^{ij}_{\alpha\beta}$ appearing on the right-hand side 
of Eq. (16) in Ref. \cite{GX}. Then one may directly obtain the explicit 
expressions of the nine-independent $S^q_{\alpha\beta}$ in terms of the 
same mixing angles and $CP$-violating phases. For illustration, 
we instructively take 
$s_{02}, s_{03}, s_{12}, s_{13} \sim \epsilon \ll 1$ \cite{Dai}. 
In this approximate but simpler case, we arrive at
\begin{eqnarray}
S_{se}^{a} & \approx & \frac{1}{2} 
\left (-c_{01} s_{01} s_{02} s_{12} \sin \phi_z - 
c_{01} s_{01} s_{03} s_{13} \sin \phi_y \right ) \; ,
\nonumber \\
S_{s\tau}^{a} & \approx & \frac{1}{2} \left [c_{23} s_{02} s_{03} s_{23} 
\sin\phi_x + c_{01} c^2_{23} s_{01} s_{03} s_{13} \sin \phi_y + c_{01}
s_{01} s_{02} s_{12} s_{23}^2 \sin \phi_z \right . 
\nonumber \\
& & \left . + c_{01} c_{23}s_{01} s_{02} s_{13} s_{23} 
\sin (\phi_x - \phi_y) - c_{01} c_{23}
s_{01} s_{03} s_{12} s_{23} \sin (\phi_x + \phi_z) \right ] \; , 
\nonumber \\
S_{e\mu}^{a} & \approx & -\frac{1}{2} \left [c_{01} c_{23}^{2} s_{01} s_{02}
s_{12} \sin \phi_z + c_{01} s_{01} s_{03} s_{13} s_{23}^{2} 
\sin\phi_y + c_{01} c_{23}
s_{01} s_{03} s_{12} s_{23} \sin (\phi_x + \phi_z) \right .
\nonumber \\
& & \left . - c_{01} c_{23}s_{01} s_{02} s_{13} s_{23} 
\sin (\phi_x - \phi_y) + c_{23}
s_{12} s_{13} s_{23} \sin (\phi_x - \phi_y + \phi_z) \right ] \; ,
\nonumber \\
S_{se}^{b} & \approx & \frac{1}{2} \left (c_{01} s_{01} s_{02} s_{12} 
\sin\phi_z - c_{01} s_{01} s_{03} s_{13} \sin \phi_y \right ) \; ,
\nonumber \\
S_{s\tau}^{b} & \approx & \frac{1}{2} \left [(c_{01}^{2}-s_{01}^{2}) c_{23}
s_{02} s_{03} s_{23} \sin \phi_x + c_{01} c^2_{23} s_{01} s_{03}
s_{13} \sin \phi_y - c_{01}
s_{01} s_{02} s_{12} s_{23}^2 \sin \phi_z \right .
\nonumber \\
& & \left . - c_{01} c_{23}s_{01} s_{02} s_{13} s_{23} 
\sin (\phi_x - \phi_y) - c_{01} c_{23}
s_{01} s_{03} s_{12} s_{23} \sin (\phi_x + \phi_z) \right ] \; ,
\nonumber \\
S_{e\mu}^{b} & \approx & -\frac{1}{2} \left [-c_{01} c_{23}^{2} s_{01} s_{02}
s_{12} \sin \phi_z + c_{01} s_{01} s_{03} s_{13} s_{23}^{2} 
\sin\phi_y - c_{01} c_{23}
s_{01} s_{03} s_{12} s_{23} \sin (\phi_x + \phi_z) \right .
\nonumber \\
& & \left . - c_{01} c_{23}s_{01} s_{02} s_{13} s_{23} \sin (\phi_x - \phi_y) 
+ (s_{01}^{2}-c_{01}^{2}) c_{23}
s_{12} s_{13} s_{23} \sin (\phi_x - \phi_y + \phi_z) \right ] \; , 
\nonumber \\
S_{se}^{c} & \approx & \frac{1}{2} \left (-c_{01} s_{01} s_{02} s_{12} 
\sin\phi_z - c_{01} s_{01} s_{03} s_{13} \sin \phi_y \right ) \; ,
\nonumber \\
S_{s\tau}^{c} & \approx & \frac{1}{2} \left [-c_{23} s_{02} s_{03} s_{23}
\sin \phi_x + c_{01} c^2_{23} s_{01} s_{03} s_{13} \sin \phi_y +
c_{01} s_{01} s_{02} s_{12} s_{23}^2 \sin \phi_z \right .
\nonumber \\
& & \left . + c_{01} c_{23}s_{01} s_{02} s_{13} s_{23} 
\sin (\phi_x - \phi_y) - c_{01} c_{23}
s_{01} s_{03} s_{12} s_{23} \sin (\phi_x + \phi_z) \right ] \; ,
\nonumber \\
S_{e\mu}^{c} & \approx & -\frac{1}{2} \left [c_{01} c_{23}^{2} s_{01} s_{02}
s_{12} \sin \phi_z + c_{01} s_{01} s_{03} s_{13} s_{23}^{2} 
\sin\phi_y + c_{01} c_{23}
s_{01} s_{03} s_{12} s_{23} \sin (\phi_x + \phi_z) \right .
\nonumber \\
& & \left . - c_{01} c_{23}s_{01} s_{02} s_{13} s_{23} 
\sin (\phi_x - \phi_y) - c_{23}
s_{12} s_{13} s_{23} \sin (\phi_x - \phi_y + \phi_z) \right ] \; ,
\end{eqnarray}
where $\phi_x \equiv \delta_{03} - \delta_{02} - \delta_{23}$,
$\phi_y \equiv \delta_{03} - \delta_{01} - \delta_{13}$ and
$\phi_z \equiv \delta_{02} - \delta_{01} - \delta_{12}$.
In obtaining these results, the corrections of 
${\cal O}(\epsilon^3)$ or smaller have been neglected. 
Note that all $S^q_{\alpha\beta}$ given in Eq. (17) 
are suppressed by the factors of ${\cal O}(\epsilon^2)$. Note also that
$S^a_{se} \approx S^c_{se}$ holds, as a consequence of 
$J^{23}_{se} \approx 0$ in the approximation made above.

As some direct relations between the rephasing invariants of $CP$ 
violation $J^{ij}_{\alpha\beta}$ and the probability asymmetries 
of neutrino oscillations $\Delta P_{\alpha\beta}$ have been
established in Ref. \cite{GX}, one can straightforwardly obtain the
relations between $\Delta_{\alpha\beta}$ and $S^q_{\alpha\beta}$
or $S^q_{ij}$ with the help of Eqs. (9) and (10). For simplicity, we do not
go into detail at this point. We shall not discuss possible terrestrial
matter effects on the unitarity quadrangles in realistic long-baseline
neutrino oscillation experiments either, because
the relevant discussions given in Refs. \cite{GX,Xing01} are
completely applicable here. Furthermore, we point out that the present 
experimental constraints on the $4\times 4$ neutrino mixing matrix
$V$ remain too poor to reveal its structural features (such as
possible symmetries or asymmetries of its off-diagonal matrix 
elements \cite{Xing02}).

In summary, we have presented a concise classification for unitarity 
quadrangles of the $4\times 4$ neutrino mixing matrix. It is found that 
there are totally thirty-six distinct topologies among twelve different
unitarity quadrangles. Useful relations between the areas of those 
unitarity quadrangles and the rephasing invariants of $CP$ violation
have been derived. For illustration, we have also expressed nine
independent areas of the unitarity quadrangles approximately
in terms of six flavor mixing angles and three $CP$-violating phases 
in the standard parametrization.

Finally, we remark that our analytical results are model-independent and 
would be very useful for a systematic study of $CP$ and $T$
violation in a variety of long-baseline neutrino oscillation experiments,
if the forthcoming MiniBooNE experiment could confirm the LSND anomaly
and support the scheme of four neutrino mixing.

\vspace{0.5cm}

This work was supported in part by the National Natural Science Foundation
of China.

\newpage

\begin{figure}
\vspace{-1cm}
\epsfig{file=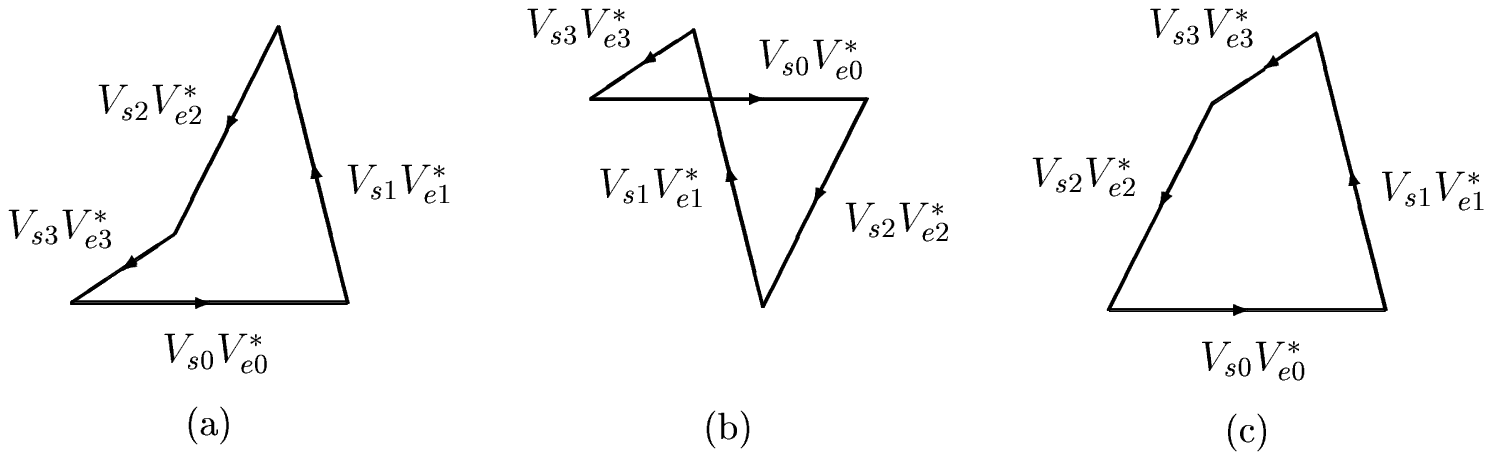,bbllx=1cm,bblly=4cm,bburx=20cm,bbury=32cm,%
width=14cm,height=22cm,angle=0,clip=}
\vspace{-13.5cm}
\caption{Three distinct topologies of unitarity quadrangle 
${\rm Q}_{se}$ .}
\end{figure}

\end{document}